\documentclass[11pt]{article}
\usepackage{epsfig}

\begin{document}

\title{Gerasimov-Drell-Hearn Sum Rule and the 
  Discrepancy between the New CLAS and SAPHIR Data}
\author{T. Mart\thanks{E-mail address: tmart@fisika.ui.ac.id}\\
Departemen Fisika, FMIPA, Universitas Indonesia, Depok, 16424, Indonesia}

\maketitle

\begin{abstract}
Contribution of the $K^+\Lambda$ channel to the Gerasimov-Drell-Hearn
(GDH) sum rule has been calculated by using the models that fit
the recent SAPHIR or CLAS differential cross section data. It is shown that the two data
sets yield quite different contributions. Contribution of this channel 
to the forward spin polarizability of the proton has been
also calculated. It is also shown that the inclusion of the recent CLAS
$C_x$ and $C_z$ data in the fitting data base does not
significantly change the result of the present calculation.
Results of the fit, however, reveal the role of the $S_{11}(1650)$, $P_{11}(1710)$, 
$P_{13}(1720)$, and $P_{13}(1900)$ resonances for 
the description of the $C_x$ and $C_z$ data. A brief discussion
on the importance of these resonances is given.
Measurements of the polarized total cross
section $\sigma_{\rm TT'}$ by the CLAS, LEPS, and MAMI collaborations 
are expected to verify this finding.\\[2ex]
\end{abstract}

\newpage
\section{\Large\sc Introduction}
The two sets of ample, good quality, experimental data of kaon
photoproduction $\gamma p\to K^+\Lambda$ 
provided recently by the SAPHIR~\cite{Glander:2003jw} 
and CLAS~\cite{Bradford:2005pt} collaborations lead to not 
only a new opportunity to study the dynamics of kaon and
hyperon interactions in a great detail, but also a puzzling point:
Why there exist discrepancies between the two data sets
in the $\gamma p\to K^+\Lambda$ channel, whereas in the
$\gamma p\to K^+\Sigma^0$ channel the extracted data from
the two collaborations agree quite well? Many recent 
efforts have been devoted to analyze the consequence of the
data discrepancies in the $\gamma p\to K^+\Lambda$ process
(see e.g., \cite{Julia-Diaz:2006is,Mart:2006dk,Bydzovsky:2006wy}). 
It is shown by Refs.~\cite{Mart:2006dk,anisovich_2005} that the use of 
SAPHIR and CLAS data, individually or simultaneously, 
leads to quite different extracted photo-coupling 
parameters. Therefore, Ref.~\cite{Mart:2006dk} concluded that
current data situation does not allow for a precise determination 
of the resonance parameters or for the search of 
the ``missing resonances''. 

By studying the statistical properties of the two data sets 
Ref.~\cite{Bydzovsky:2006wy} showed that the CLAS data are 
in good agreement with the LEPS data~\cite{Sumihama:2005er},
whereas the SAPHIR data are coherently shifted down with respect 
to both CLAS and LEPS data, especially at forward kaon angles.
The relative-global-scaling factor between the SAPHIR and CLAS 
data has been found to be 1.13. 

Although results of the recent works could reveal certain 
consequences of using SAPHIR or CLAS data in the database, 
it is still difficult to determine which data set should be 
used to obtain a reliable phenomenological model as well as
to extract the correct resonance parameters. The reason 
is that in all analyses the experimental data are
fitted by adjusting a set of free parameters, while 
the precise values of these parameters are not well known.
Furthermore, the extracted parameters are not unique and
also sensitive to the number of resonances used in a model.
Here, it is important to note that due to the high-energy
threshold it is difficult to identify which resonances
are dominant in this reaction.

In view of this, it is important to consider other quantities which
can be predicted by the models and can be directly
compared with the results from other measurements or 
model predictions. One of the possible quantities
is the contribution of the $\gamma p\to K^+\Lambda$ channel
to the Gerasimov-Drell-Hearn (GDH) sum rule. Previous 
calculations based on isobar 
models~\cite{hammer:1997,Sumowidagdo:1999yp,kamalov00,mart08}  
have estimated that kaon photoproductions on the proton 
contribute about $+4~\mu$b to the GDH integral. Actually,
to arrive at a correct GDH sum rule prediction one merely needs
about $+2$ $\mu$b contribution from kaon-hyperon final states (see
Table I of Ref.~\cite{kamalov00}).

Recently, the CLAS collaboration released a set of experimental 
data on the beam-recoil polarization observables $C_x$ and 
$C_z$~\cite{Bradford:2006ba}.
These data indicate that the $\Lambda$ polarization is 
predominantly in the direction of the spin of the 
incoming photon, independent of the c.m. energy or 
the kaon scattering angle. 
By using a circularly polarized photon beam the polarization
of the recoiling hyperon is measured. The measured polarization
is defined through the relation
\begin{equation}
  R_\Lambda = \sqrt{P_\Lambda^2+C_x^2+C_z^2}~,
\label{eq:polarization}
\end{equation}
where $P_\Lambda$ is the induced (recoil) polarization of the hyperon.
We note that for a circularly polarized beam this polarization is
bounded to be less than or equal to one~\cite{Artru:2006xf}. 
When integrated over all production angles
and total c.m. energies, the CLAS data gives
\begin{eqnarray*}
  R_\Lambda = 1.01\pm 0.01~.
\end{eqnarray*}
However, this result seems to be difficult to explain. 
Reference~\cite{Schumacher:2006ii} tried to explain this
phenomenon by means of a simple quark model. Within this
approach, the real photon converts to an $s\bar{s}$ pair 
as part of the interaction in the complex gluon field of the nucleon. 
The pair carries the polarization of the photon. The s quark
of the pair merges with the $ud$ quarks of the proton to form
a $\Lambda$. The s quark in the $\Lambda$ retains its full 
polarization after being precessed by a spin-orbit interaction, 
while the $\bar{s}$ quark merges with the remnant $u$ quark and ends 
up in the spinless kaon. Using this model Ref.~\cite{Schumacher:2006ii}
is able to qualitatively explain the phenomenon. However, as pointed
out by Ref.~\cite{Anisovich:2007bq}, in this approach the $s$-channel 
baryon resonances cannot play an important role, since the 
$s\bar{s}$ pair is created in the initial state. This is in contrary
to the fact that kaon photoproduction is dominated by the resonance
contributions. Moreover, the approach is unable to reproduce the
qualitative features of the  $\gamma p\to K^+\Sigma^0$  
process~\cite{Anisovich:2007bq}. 
Based on these facts, it is certainly interesting to include
the CLAS $C_x$ and $C_z$ data in the present analysis and to
investigate the effects of the data inclusion.

This paper is organized as follows. In Section \ref{gdh-sum-rule} 
we present the formalism of the GDH sum rule. Section~\ref{multipole-model}
briefly explains the multipole model used in the data analysis. 
Section~\ref{contribution-to-GDH} presents the results before
we include the CLAS $C_x$ and $C_z$ data in our analysis. 
The effect of the inclusion of these data will be demonstrated in
Section~\ref{after_incl}. We will summarize our findings in
Section~\ref{conclusion}.

\section{\Large\sc The GDH sum rule}
\label{gdh-sum-rule}

The GDH sum rule \cite{gdh} relates the anomalous magnetic 
moment of the nucleon $\kappa_N$  to the difference of its
polarized total photoabsorption cross sections. For the case
of proton the sum rule reads
\begin{eqnarray}
-\frac{\kappa_p^2}{4} = \frac{m_p^2}{8\pi^2\alpha}
\int_{E_\gamma^{\rm thr}}^{\infty}\frac{dE_\gamma}{E_\gamma}\,
\left[\,\sigma_{1/2}(E_\gamma)-\sigma_{3/2}(E_\gamma)\,\right]\, ,
\label{eq:igdh0}
\end{eqnarray}
where $\sigma_{3/2}$ and $\sigma_{1/2}$ indicate the cross sections 
for the possible combinations of spins of 
the proton (1/2) and the photon (1), 
$E_\gamma^{\rm thr}$ the photoproduction threshold lab energy,
$\alpha=e^2/4\pi=1/137$ the fine structure constant, and $m_p$ 
the proton mass. 

For unpolarized photoproduction experiments the total cross section 
$\sigma_{\rm T}$ can be related to the spin dependent 
cross sections by 
\begin{eqnarray}
\sigma_{\rm T} = \frac{1}{2}\,(\sigma_{3/2} + \sigma_{1/2}) ~,
\label{sigtot}
\end{eqnarray}
while for the photoproduction from the polarized
photon beam and polarized proton target, 
\begin{eqnarray}
\vec{\gamma} + \vec{p} &\rightarrow& K^+ + \Lambda \, ~,
\label{reaction}
\end{eqnarray}
one can also measure 
\begin{eqnarray}
\sigma_{\rm TT'} = \frac{1}{2}\, (\sigma_{3/2}-\sigma_{1/2})~.
\label{sigtt}
\end{eqnarray}
The latter is obviously the suitable observable for the sum rule. However,
since there has been no available measurement of the reaction given
in Eq.~(\ref{reaction}) yet, Eq.~(\ref{sigtt}) should be estimated 
from a reliable model which fits all
available unpolarized experimental data.

For the sake of comparison, we follow the notation of 
Ref.\,\cite{kamalov00} and define the GDH integral 
\begin{eqnarray}
I_{\rm GDH} \equiv
\int_{E_\gamma^{\rm thr}}^{\infty}\frac{dE_\gamma}{E_\gamma}\,
\left[\,\sigma_{1/2}(E_\gamma)-\sigma_{3/2}(E_\gamma)\,\right]\, .
\label{eq:igdh}
\end{eqnarray}
With this definition, the GDH sum rule for the proton yields 
\begin{eqnarray}
I_{\rm GDH} =  -\frac{2\pi^2\alpha\kappa^2}{m_p^2} = -204.5~\mu{\rm b}\, .
\label{eq:igdh1}
\end{eqnarray}
The first estimate based on the then existing data led to $I_{\rm GDH}=-261~\mu{\rm b}$
\cite{karliner73}. The latest result calculated from a combination of pion photoproduction
and photon absorption processes yields a value of $-211\pm 15$ $\mu$b \cite{Drechsel:2007sq},
thus, although it slightly overestimates the sum rule, it is still consistent
within the present error bar. However, one can also calculate the $I_{\rm GDH}$
by summing up all possible photoproduction processes. This was done in Ref.~\cite{kamalov00},
and it is found that in order to arrive at a consistent result, an estimate of +4 $\mu$b 
contribution should come from kaon photoproduction processes on the proton. By using
KAON-MAID it can be shown that a slightly smaller value of $1.25+1.38+0.30=2.93$ $\mu$b
can be expected from these associated strangeness processes, where the three separate
contributions refer to the $K^+\Lambda$, $K^+\Sigma^0$, and $K^0\Sigma^+$ channels
\cite{mart08}. The small contribution from the $K^0\Sigma^+$ channel can be understood
from the fact that the cross section of this channel is substantially smaller than
those of the $K^+\Lambda$ and $K^+\Sigma^0$ channels (see, e.g., Ref.~\cite{Lawall:2005np}).
The latest measurement of the $K^0\Sigma^+$ channel~\cite{Castelijns:2007qt} shows
that the corresponding cross section is even smaller than that
reported in Ref.~\cite{Lawall:2005np}. This indicates that the
contribution of the kaon-hyperon final states to the GDH sum rule
for the case of the proton probably comes only from the
$K^+\Lambda$ and $K^+\Sigma^0$ channels.

With the knowledge of $\sigma_{\rm TT'}$ we can also calculate the 
corresponding contribution to the forward spin polarizability of the proton,
\begin{equation}
\gamma_0 = \frac{1}{4\pi^2} 
\int_{E_\gamma^{\rm thr}}^{\infty}\frac{dE_\gamma}{E_\gamma^3}\,
\left[\,\sigma_{1/2}(E_\gamma)-\sigma_{3/2}(E_\gamma)\,\right]\, .
\label{eq:forward_polar}
\end{equation}
The precise value of this observable is currently less known, 
since there is no sign that calculations from Chiral Perturbation Theory 
(ChPT) would converge in this case. For instance, 
the isobar model of Ref.\,\cite{kamalov00} predicts
$\gamma_0 = -0.65$ (in $10^{-4}$ fm$^4$), whereas the analysis
from ChPT to order ${\cal O}(p^3)$ obtains a value of $4.6$
\cite{bkm95} and an extension to ${\cal O}(p^4)$ yields $-3.9$
\cite{jko+kmb}. The latest value obtained from photoabsorption
experiment \cite{Drechsel:2007sq} is $-0.94\times 10^{-4}$ fm$^4$.
In spite of this enormous uncertainty it is important to note
that Ref.~\cite{kamalov00} estimates a small contribution from
the $n\pi+\eta$ channels, i.e. $-0.01\times 10^{-4}$ fm$^4$,
in spite of the fact that the $n\pi$ channel could have total a cross
section up to five orders of magnitude larger than that of kaon
channels.

\section{\Large\sc The multipole model}
\label{multipole-model}
One of the most recent models of the $K^+\Lambda$ photoproduction 
which fits all recent-experimental data, in a wide range of kinematics, 
from threshold up to $E_\gamma\approx 3$ GeV
is the recent multipole model given in Ref.~\cite{Mart:2006dk}. 
The model consists of the background terms which are constructed from a series 
of tree-level Feynman diagrams and the resonance terms which are 
assumed to have the Breit-Wigner form. To account for hadronic structures 
of interacting baryons and mesons the appropriate hadronic form 
factors are included in hadronic vertices in a gauge-invariant
fashion. Details of the ingredient of the model can be found in 
Ref.~\cite{Mart:2006dk}. A number of free parameters in the background
and resonance terms are adjusted by fitting the calculated observables to
experimental data. Due to the problem of the data discrepancies,
two models have been proposed in Ref.~\cite{Mart:2006dk}. The first
model fits the combination of the SAPHIR and LEPS data (hereafter 
referred to as Fit 1a) and the second one fits the combination of 
the CLAS and LEPS data (hereafter referred to as Fit 2a). Before
the inclusion of the CLAS beam-recoil polarization data $C_x$ and $C_z$, 
in total, 834 data points were fitted in the first case and 
1694 data points were fitted in the second case.
To further analyze the physical consequence of the data discrepancy, 
we have refitted this multipole model solely to the SAPHIR data 
(750 data points) or CLAS data (1377 data points). 
The results are referred to as Fit 1 (fits to the SAPHIR data) 
and Fit 2 (fits to the CLAS data). The CLAS beam-recoil polarization 
data $C_x$ and $C_z$ consists of 320 data points.
The results after the inclusion 
of these data will be presented in Section~\ref{after_incl}.

\section{\Large\sc Contribution to the GDH sum rule}
\label{contribution-to-GDH}
\begin{figure}[!h]
  \begin{center}
    \leavevmode
    \epsfig{figure=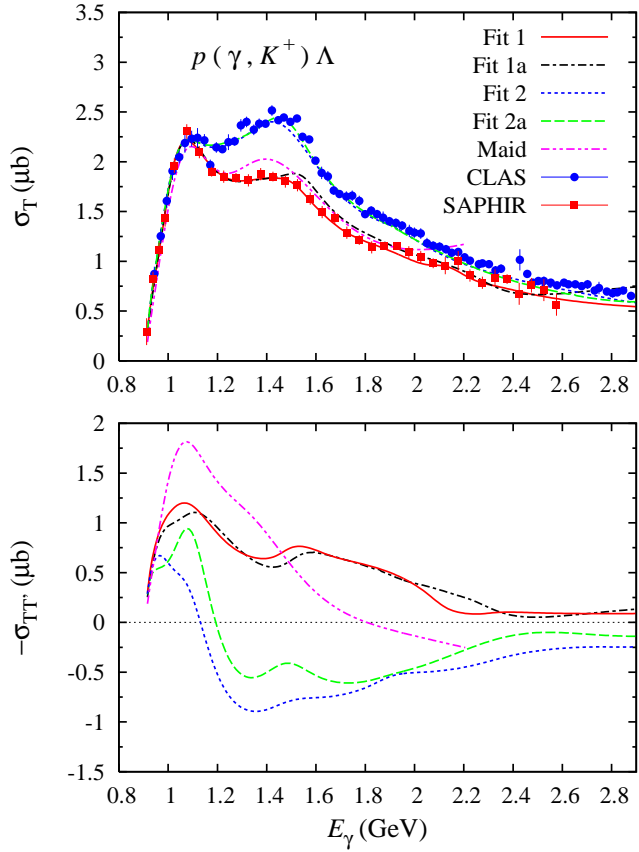,width=90mm}
    \caption{(Color online) Total cross sections $\sigma_{T}$ (upper panel)
    and $-\sigma_{TT'}$ (lower panel) for the $\gamma p\to K^+\Lambda$ 
    channel plotted as a function of the photon laboratory energy $E_\gamma$. 
    Fit 1a and Fit 2a show the result of a multipole analysis
    of Ref. \protect\cite{Mart:2006dk} that fits SAPHIR and CLAS data,
    respectively. Fit 1 and Fit 2 demonstrate the result of the
    same multipole analysis, but refitted solely to the SAPHIR 
    and CLAS data, respectively. The prediction of the KAON-MAID model is
    indicated in both panels by Maid.
    Solid squares and solid circles display experimental 
    data from the SAPHIR \protect\cite{Glander:2003jw} and the 
    CLAS \protect\cite{Bradford:2005pt} collaborations, respectively.
    Note that all data shown in this figure were not used in the fits.}
   \label{fig:total} 
  \end{center}
\end{figure}

Figure~\ref{fig:total} displays the comparison between the predicted
total cross sections $\sigma_{\rm T}$ and experimental data in the
upper panel. Predictions for the $\sigma_{\rm TT'}$ total cross section
are given in the lower panel of the same figure. 
For the sake of completeness, we also present the prediction
of KAON-MAID (indicated by Maid in the figure).

Since the number of experimental data used in the new fits 
(Fit 1 and Fit 2) is smaller, it is 
understandable that the agreement between experimental data and model 
predictions is better in this case. The predictions for the $\sigma_{\rm TT'}$ 
show, however, quite different behaviors for different models. 
The differences between Fit 1 and Fit 1a (as well as between 
Fit 2 and Fit 2a) originate from the differences in the fitted (resonance) parameters,
thus, reflecting the differences in the resonance roles in explaining the process.
Nevertheless, the gross behaviors of Fit 1 and Fit 1a are almost the same. 
The same feature is also displayed by Fit 2 and Fit 2a. The main difference is,
whereas  Fit 1 and Fit 1a do not cross the zero axis, Fit 2 and Fit 2a change 
their signs at $E_\gamma\approx 1.2$ GeV. Based on this result 
we can then expect smaller $I_{\rm GDH}$ values in the case of
Fit 2 and Fit 2a (the case of using CLAS data).

\smallskip

\begin{table}
\caption{Contributions of  the $\gamma p\to K^+\Lambda$ channel 
         to the GDH sum rule for the proton in $\mu$b and to the 
         forward spin polarizability $\gamma_0$ in $10^{-7}~$ fm$^4$. 
         Note that the numerical values of the $I_{\rm GDH}$ refer
         to $\int_{E_\gamma^{\rm thr}}^{\infty}dE_\gamma\,(\sigma_{1/2}-\sigma_{3/2})/E_\gamma$
         in Eq.\,(\ref{eq:igdh}). For comparison, the GDH sum rule
         yields the value of $-204.5$ $\mu$b, while Ref.\,\protect\cite{kamalov00}
         estimates a value of $+4$ $\mu$b for all kaon photoproduction channels
         on the proton. On the other hand, Ref.\,\protect\cite{kamalov00} 
         obtains a value of $-0.65\times 10^{-4}$ fm$^4$
         for the contribution from single pion photoproduction 
         below 1.6 GeV to the $\gamma_0$. The number of fitted
         data $N$ and the corresponding $\chi^2$ per degrees of freedom
         are also given.\label{tab:num_res}}
\begin{center}
\begin{tabular}{lrrrrr}
\hline\hline
Observable& ~~~~~~~MAID& ~~~~~Fit 1 & ~~~~~Fit 1a & ~~~~~Fit 2 & ~~~~~Fit 2a  \\
\hline
    $I_{\rm GDH}$ ($\mu$b) & 1.247&1.309 & 1.274 & $-0.845$ & $-0.333$ \\
    $\gamma_0$ ($10^{-7}$ fm$^4$)~~~~~ & 0.939& 0.807 & 0.753 & $-0.208$ & 0.060 \\
    $N$ & 319& 720& 834 & 1377& 1694 \\
    $\chi^2/N_{\rm dof}$ & 3.36& 0.78 & 1.02 & 0.84 & 0.98 \\
\hline\hline
\end{tabular}
\end{center}
\end{table}

\begin{figure}[!h]
  \begin{center}
    \leavevmode
    \epsfig{figure=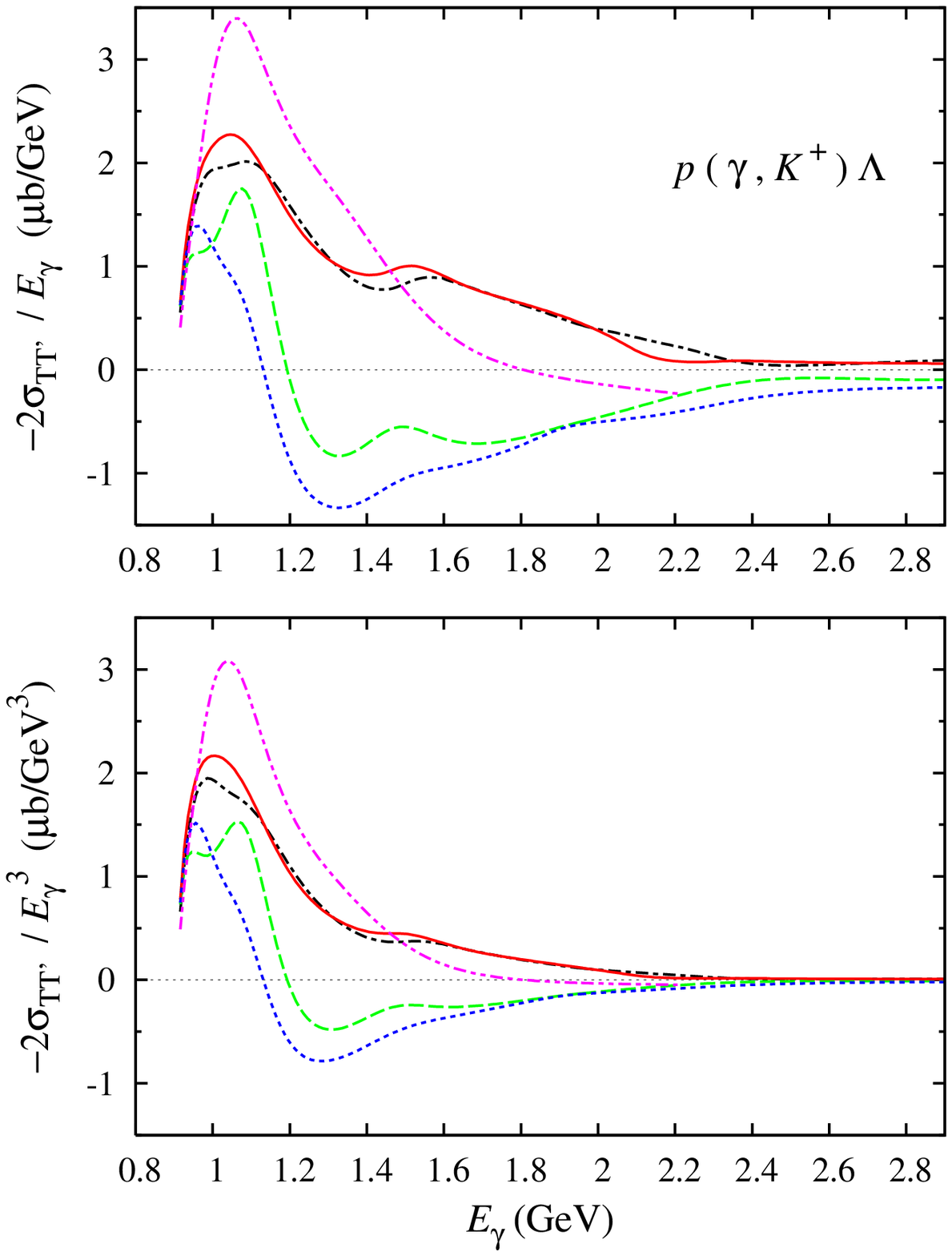,width=90mm}
    \caption{(Color online) The integrands of 
    Eq.~(\ref{eq:igdh}) (upper panel)
    and Eq.~(\ref{eq:forward_polar}) (lower panel) for all five models 
    given in Table~\protect\ref{tab:num_res} as a function
    of the photon laboratory energy $E_\gamma$. Notation for the curves
    is as in Fig.~\ref{fig:total}.}
   \label{fig:integrand} 
  \end{center}
\end{figure}

The negative and small values of $\gamma_0$ predicted by Fit 2 and Fit 2a,
respectively, are also expected, since the integrands of Eqs.~(\ref{eq:igdh}) 
and (\ref{eq:forward_polar}) are similar up to the power in the 
denominator. 

The numerical results of the $I_{\rm GDH}$ shown in  Table~\ref{tab:num_res}
prove our expectations. To help to understand these numerical 
values we present the evolutions of the integrands of Eqs.~(\ref{eq:igdh}) 
and (\ref{eq:forward_polar}) in Fig.~\ref{fig:integrand}. The predicted
$I_{\rm GDH}$ of Fit 1 and Fit 1a are much closer to the MAID prediction
than those of Fit 2 and Fit 2a, indicating the consistency of the new
SAPHIR data \cite{Glander:2003jw} to the old ones \cite{saphir98}. The 
origin of the negative values obtained from Fit 2 and Fit 2a is obvious
from the evolution of both integrands shown by the dotted and dashed
curves in the upper panel of Fig.~\ref{fig:integrand}. 

It comes as no surprise that 
the predicted values of $\gamma_0$ are quite small compared to the presently
known value, i.e., $-0.94\times 10^{-4}$ fm$^4$  \cite{Drechsel:2007sq}.
However, it is interesting to compare the contribution from  
the $n\pi+\eta$ channels, i.e. $-1\times 10^{-6}$ fm$^4$, to those
predicted by both Fit 1 and Fit 1a (as well as MAID), i.e.  
$\gamma_0\approx +1\times 10^{-7}$ fm$^4$. The absolute value of the 
latter is one order of magnitude smaller than the former, although
the total cross section $\sigma_{\rm T}$ of the latter could be
five orders of magnitude smaller than the former. This result
recommends a further analysis of the contribution from  
the $n\pi+\eta$ channels to the $\gamma_0$. 

The different behaviors shown by Fit 2 and Fit 2a can be traced
back to the role of the background and resonance terms in these
models. It is shown in Ref.~\cite{Mart:2006dk} that, in contrast
to the model that fits the SAPHIR data, model that fits the
CLAS data (in this case Fit 2a)
yields an unrealistically large background contribution.
To overcome this, the resonance contributions should produce 
a kind of destructive interference with the background terms.
Another problem with the Fit 2a model is shown in 
Ref.~\cite{Mart:2007mp}. In spite of the fact that Fit 2a was
fitted to the CLAS photoproduction data, it fails to reproduce
the CLAS electroproduction data~\cite{Ambrozewicz:2006zj}.
Surprisingly, this is in contrast to the Fit 1a model, which
is fitted to the SAPHIR photoproduction data. 

\section{\Large\sc Inclusion of the $C_x$ and $C_z$ data}
\label{after_incl}
Traditionally, the beam-recoil polarization observables are calculated
in the system where the $z'$ is defined by the direction of the outgoing
kaon \cite{knoechlein}. The observables are therefore called by $C_{x'}$ and $C_{z'}$. 
The CLAS collaboration measured these observables, however, in the
system where the $z$ axis of the reaction plane is along the direction
of the photon beam~\cite{Bradford:2006ba}. Therefore, in our calculation 
we should consider $C_{x}$ and $C_{z}$ which can be obtained 
from the standard rotation matrix, 
\begin{eqnarray}
  \label{eq:rotation}
  C_x &=& C_{x'}\cos\theta+C_{z'}\sin\theta ~,\\
  C_z &=& -C_{x'}\sin\theta+C_{z'}\cos\theta~,
\end{eqnarray}
where $\theta$ is the kaon scattering angle. For the definition of
$C_{x'}$ and $C_{z'}$ we follow Refs.~\cite{Bradford:2006ba,Anisovich:2007bq}.

\begin{figure}[!t]
  \begin{center}
    \leavevmode
    \epsfig{figure=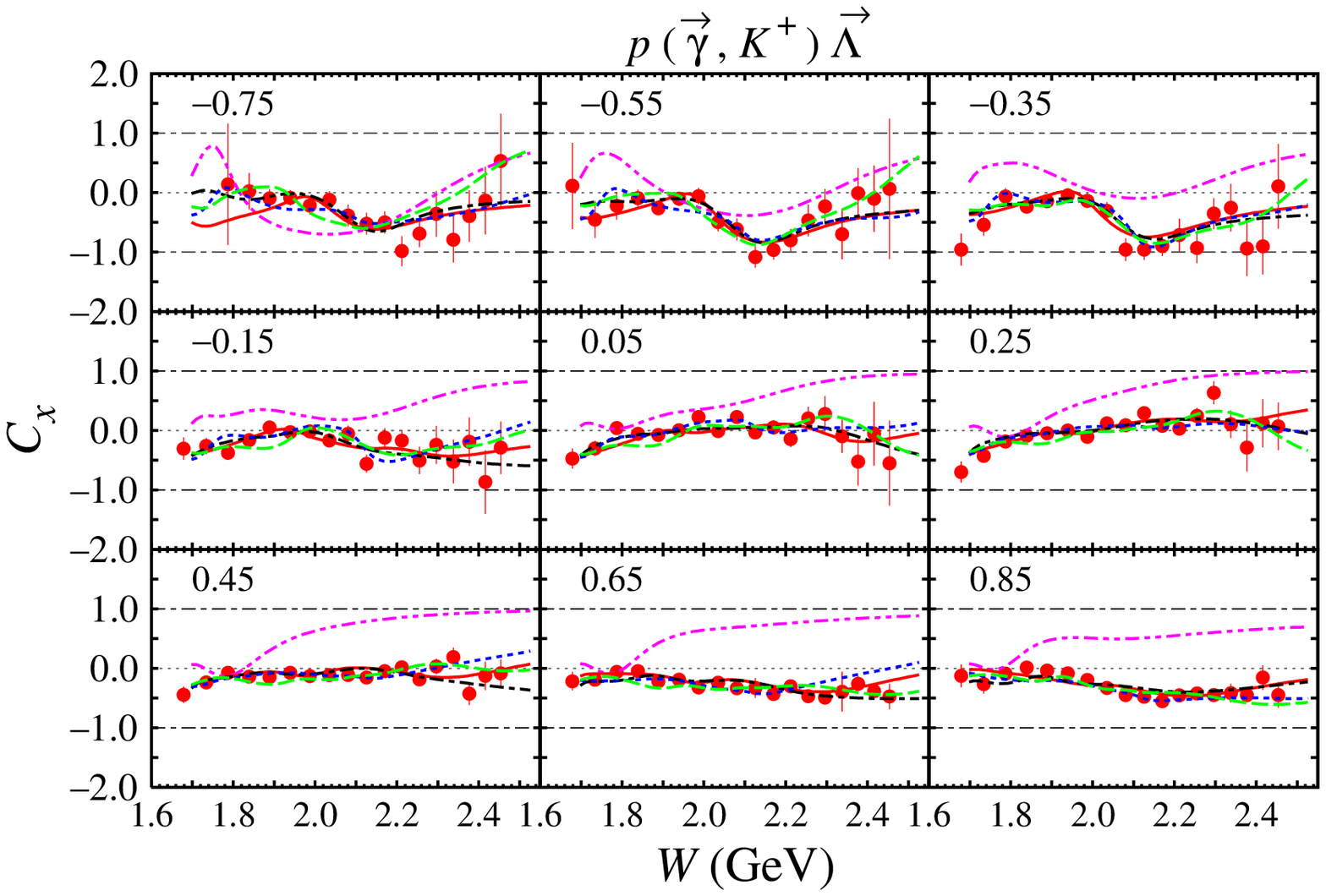,width=110mm}
    \caption{(Color online) The beam-recoil observable $C_x$ 
      for the reaction $\vec{\gamma} p\to K^+\vec{\Lambda}$ 
      plotted as a function of the total c.m. energy $W$. Experimental data
      are taken from Ref.~\cite{Bradford:2006ba}. The corresponding value 
      of $\cos\theta$ is shown in each panel. Notation of the curves
      is as in Fig.~\ref{fig:total}.}
   \label{fig:cx_e} 
  \end{center}
\end{figure}

\begin{figure}[!]
  \begin{center}
    \leavevmode
    \epsfig{figure=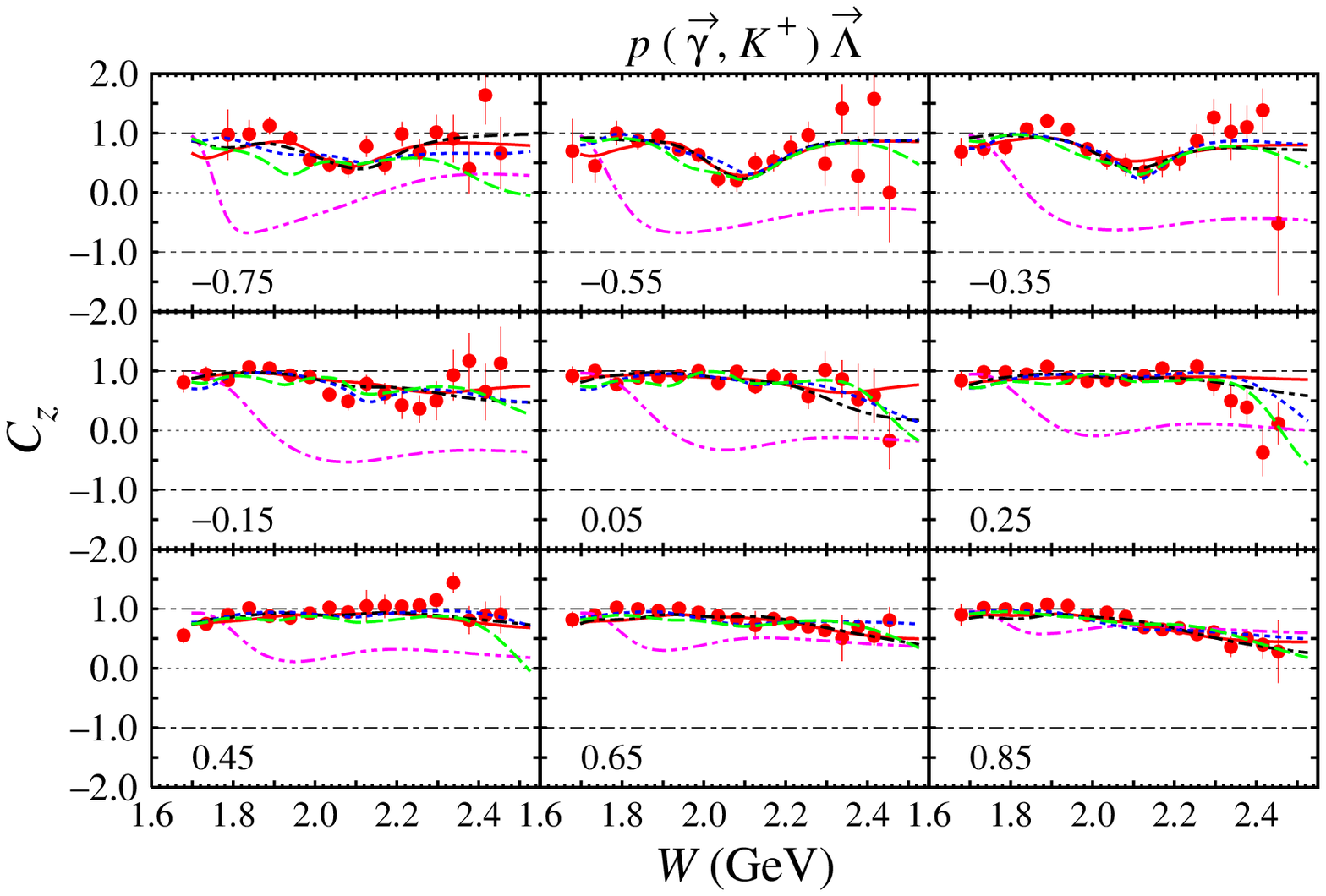,width=110mm}
    \caption{(Color online) As in Fig.~\ref{fig:cx_e}, but for the
      observable $C_z$.}
   \label{fig:cz_e} 
  \end{center}
\end{figure}

\begin{figure}[!h]
  \begin{center}
    \leavevmode
    \epsfig{figure=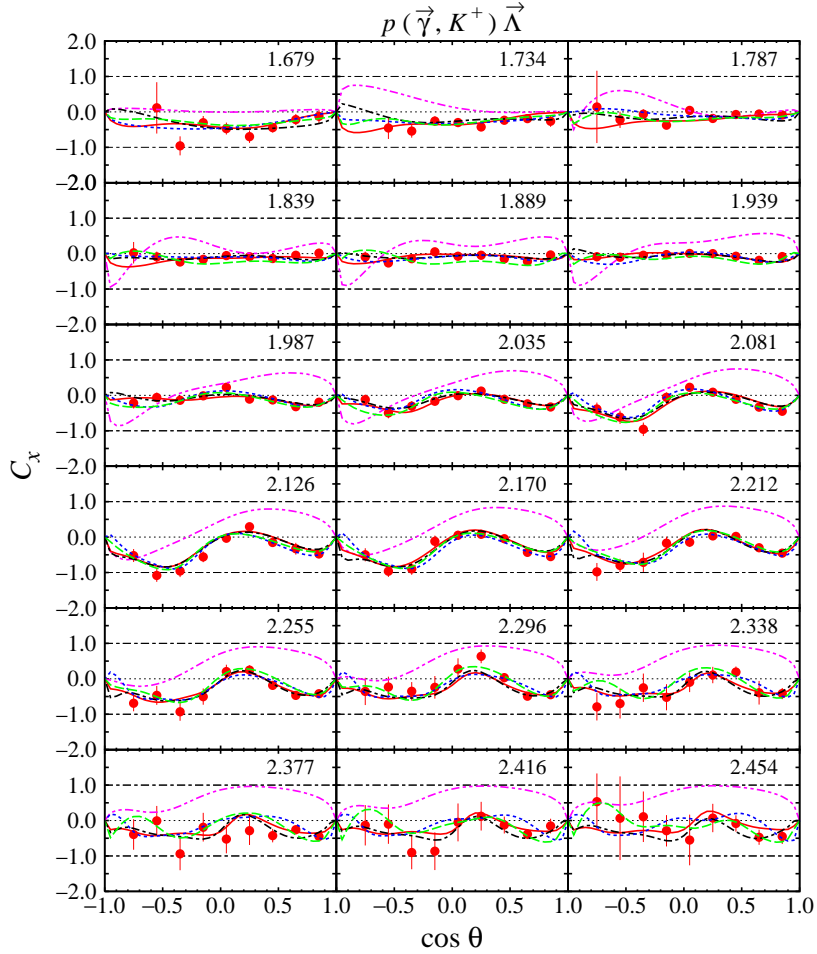,width=110mm}
    \caption{(Color online) The beam-recoil observable $C_x$ plotted 
      as a function of the kaon scattering angle. Notations of the curves
      and experimental data are as in Fig.~\ref{fig:cx_e}. Shown in each
      panel is the total c.m. energy $W$ in GeV.}
   \label{fig:cx_th} 
  \end{center}
\end{figure}

\begin{figure}[!h]
  \begin{center}
    \leavevmode
    \epsfig{figure=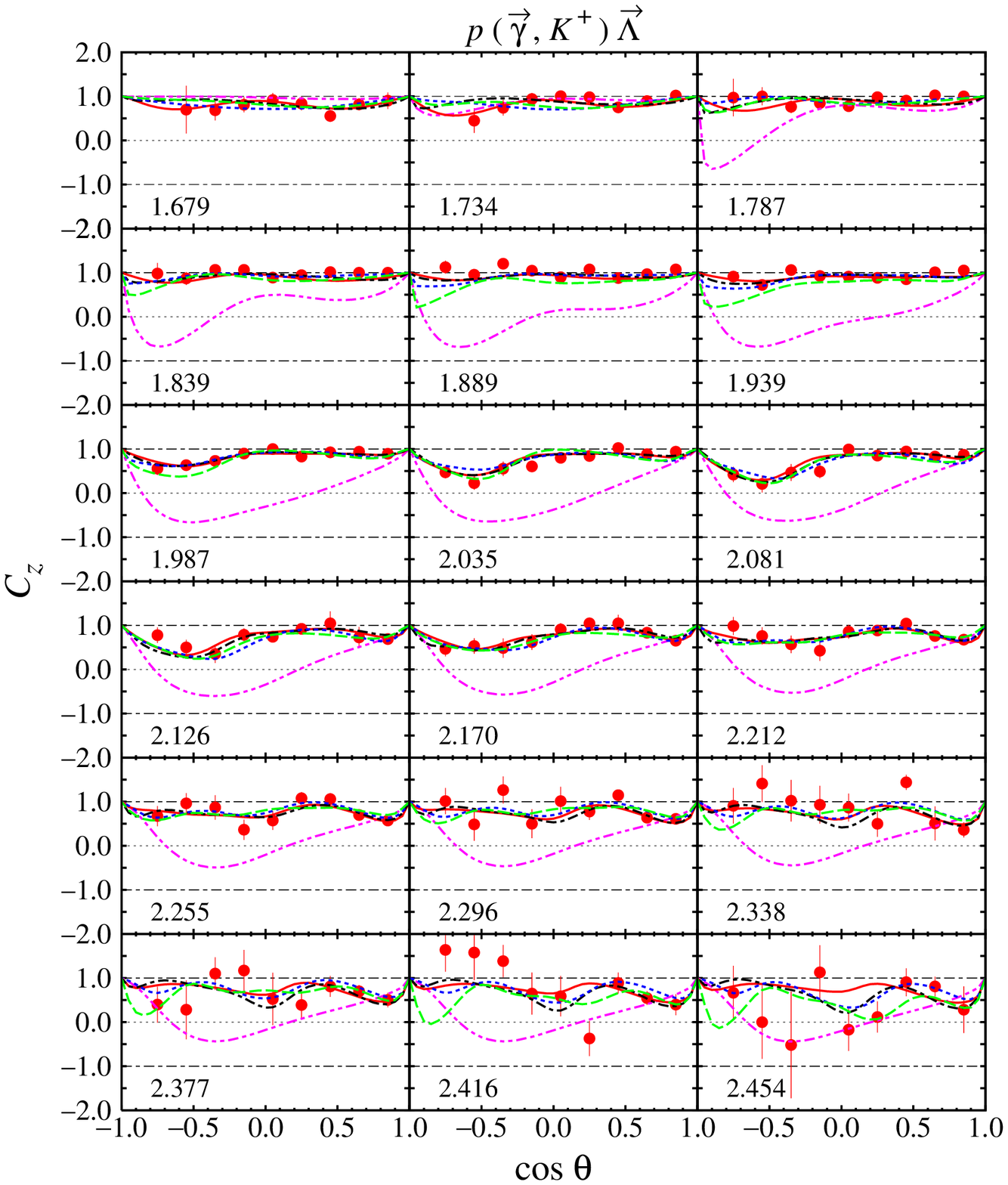,width=110mm}
    \caption{(Color online) As in Fig.~\ref{fig:cx_th}, but for the
      observable $C_z$.}
   \label{fig:cz_th} 
  \end{center}
\end{figure}

After including the CLAS $C_x$ and $C_z$ data in the fitting database
we refit the Fit 1, Fit 1a, Fit 2, and Fit 2a models and denote the
refitted models by Fit 1x, Fit 1ax, Fit 2x, and Fit 2ax. As shown
in Figs.~\ref{fig:cx_e}--\ref{fig:cz_th} the agreement between 
the calculated and the experimental data of $C_x$ and $C_z$
can be satisfactorily achieved. We also find that in order to fit 
these data we do not need a weighting factor as in 
Ref.~\cite{Anisovich:2007bq}. This is clearly demonstrated by 
the agreement of the calculated and experimental data of $C_x$ and $C_z$
as well as the $\chi^2/N_{\rm dof}$ listed in Table~\ref{tab:num_resx},
where in all cases the latter increases only slightly. At higher $W$ all models
start to vary, because experimental data at this kinematics have large
error bars. The disagreement of KAON-MAID with the recent CLAS $C_x$ 
and $C_z$ data is expected, since this model has very few nucleon 
resonances and was fitted to a small number of experimental data 

\begin{figure}[!h]
  \begin{center}
    \leavevmode
    \epsfig{figure=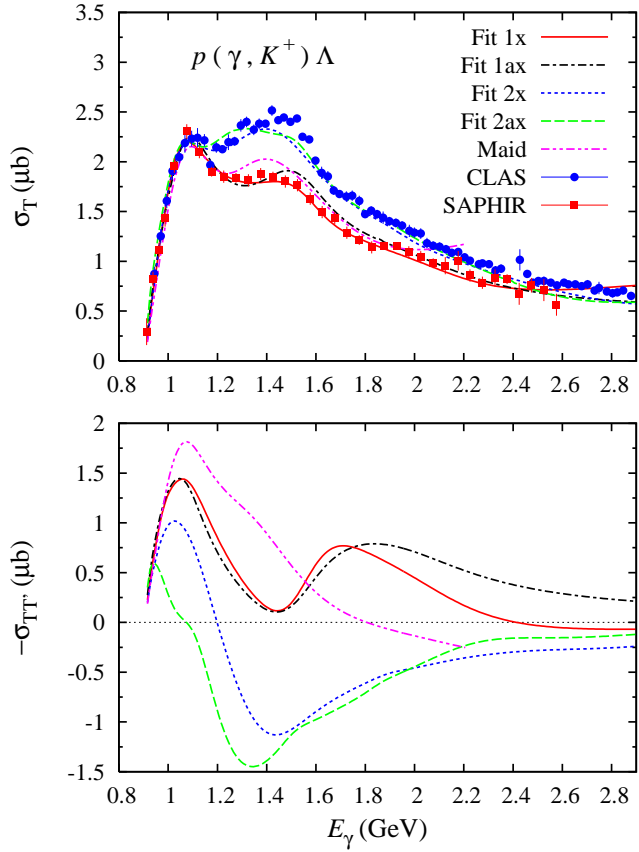,width=90mm}
    \caption{(Color online) As in Fig.~\ref{fig:total}, but all curves (except those
      of Kaon-Maid) are obtained from the same models which have been refitted 
      after the inclusion
      of the beam-recoil polarization data $C_x$ and $C_z$ in the fitting database.}
   \label{fig:sigttp_cxcz} 
  \end{center}
\end{figure}

\begin{figure}[!h]
  \begin{center}
    \leavevmode
    \epsfig{figure=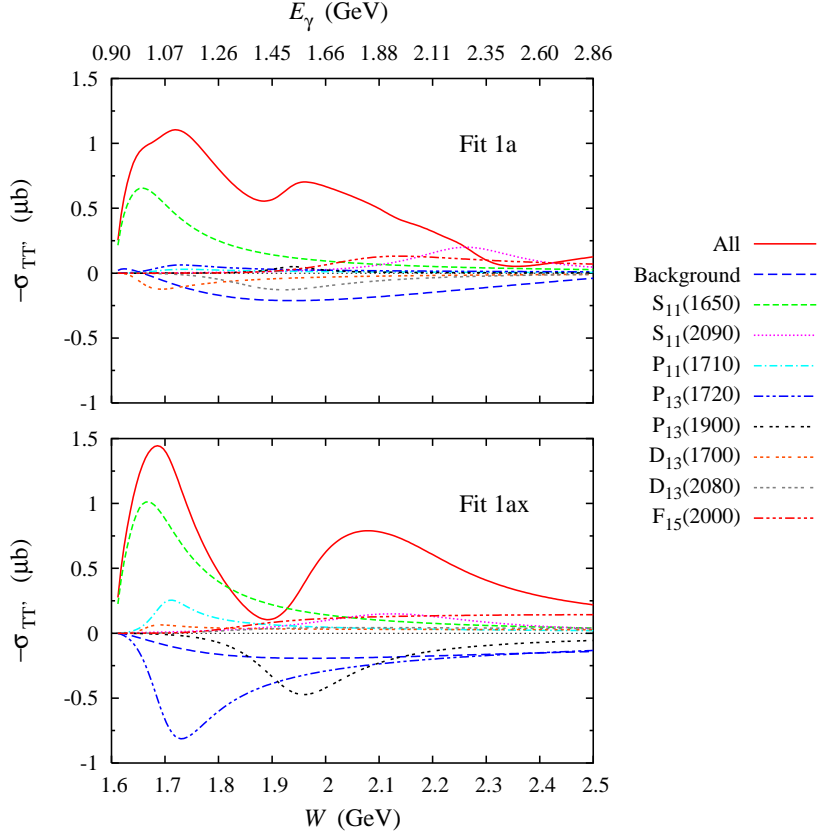,width=110mm}
    \caption{(Color online) Contribution of the background terms and several 
      important resonances to the total cross section $\sigma_{\rm TT'}$ 
      before (Fit 1a) and after (Fit 1ax) the inclusion of the beam-recoil
      polarization observables $C_x$ and $C_z$. The total cross sections 
      $\sigma_{\rm TT'}$ obtained by including all terms are indicated by
      the solid lines. Note that contributions from other resonances 
      are small and, therefore, are not shown in this figure for the sake of clarity.}
   \label{fig:res_cont} 
  \end{center}
\end{figure}

\begin{table}
\caption{As in Table~\protect\ref{tab:num_res}, but obtained
  after the inclusion of the $C_x$ and $C_z$ data. 
  Note that the result of KAON-MAID has been excluded from this Table.\label{tab:num_resx}}
\begin{center}
\begin{tabular}{lrrrrr}
\hline\hline
Observable&  ~~~~~Fit 1x & ~~~~~Fit 1ax & ~~~~~Fit 2x & ~~~~~Fit 2ax  \\
\hline
    $I_{\rm GDH}$ ($\mu$b)            & 1.140 & 1.380 & $-0.642$ & $-1.181$ \\
    $\gamma_0$ ($10^{-7}$ fm$^4$)~~~~~& 0.752 & 0.785 & $-0.033$ &$-0.466$ \\
    $N$                               & 1040  & 1154  & 1697     & 2014 \\
    $\chi^2/N_{\rm dof}$              & 0.93  & 1.13  & 0.91     & 1.28 \\
\hline\hline
\end{tabular}
\end{center}
\end{table}

The total cross sections $\sigma_{\rm T}$ and $\sigma_{\rm TT'}$ of the
refitted models display an interesting result. As shown in Fig.~\ref{fig:sigttp_cxcz},
the inclusion of the $C_x$ and $C_z$ data emphasizes our previous finding;
the models which fitted the CLAS differential cross section data predicts 
a (more) negative contribution to the GDH sum rule. The numerical results given 
in Table~\ref{tab:num_resx} show this explicitely. Given that there is
no discrepancy between the SAPHIR and CLAS $K^+\Sigma^0$ data, so that
the previous result of the $K^+\Sigma^0$ channel is still valid, and
the contribution of the $K^0\Sigma^+$ channel can be neglected, we can
conclude that models which fit the CLAS differential cross section data
(Fit 2x and Fit 2ax) tend to eliminate the contribution of kaon-hyperon
final states to the GDH sum rule. 

It is also important to briefly discuss
the differences between the predicted $\sigma_{\rm TT'}$ before and
after the inclusion of the $C_x$ and $C_z$ data (shown in the lower
panels of Figs.~\ref{fig:total} and \ref{fig:sigttp_cxcz}). After the
inclusion of the beam-recoil data the total cross sections 
$\sigma_{\rm TT'}$ show
less structures. Presumably, the CLAS $C_x$ and $C_z$ data select
certain resonances as the important ones. To investigate this, in
Fig.~\ref{fig:res_cont} we plot contributions of the background 
terms and several important resonances to the total cross section 
$\sigma_{\rm TT'}$ before and after the inclusion of the beam-recoil
observables. Although the phenomenon is observed in all four fits,
we only compare Fit 1a and Fit 1ax because the 
effect of the $C_x$ and $C_z$ inclusion is more dramatic in this case.

By comparing the upper and lower panels of 
Fig.~\ref{fig:res_cont} we can conclude that the 
inclusion of the $C_x$ and 
$C_z$ data does not influence the background sector of the model.
In the resonance sector, this inclusion emphasizes the roles of
the $S_{11}(1650)$, $P_{11}(1710)$, $P_{13}(1720)$, and $P_{13}(1900)$
resonances. The result of the present work, therefore, corroborates
the finding of the authors of Ref.~\cite{Anisovich:2007bq}
about the evidence of the $P_{13}(1900)$ resonance, who also used 
the CLAS $C_x$ and $C_z$ data in their analysis. We note that this
analysis has been further reexamined in a great detail in 
Ref.~\cite{Nikonov:2007br}, where it is shown that 
the $P_{13}(1900)$ resonance is essential to achieve a good
quality of the fit, especially for the $C_x$ and $C_z$ data. 
It is also important to note that Ref.~\cite{Mart:2006dk} 
has pointed out that this resonance is important if the CLAS 
differential cross section data were used. 
With the same quantum states, another important resonance is the 
$P_{13}(1720)$. Reference~\cite{Nikonov:2007br} also found that this 
resonance belongs to the four strongest contributors 
to the $\gamma p \to K^+\Lambda$ reaction. The importance of
the $S_{11}(1650)$ in the photoproduction of $K^+\Lambda$ 
is quite clear because the $s$-wave contribution dominates
the threshold of this process~\cite{saphir98}. This has 
been also discussed in the previous 
works~\cite{Glander:2003jw,Mart:2006dk,Anisovich:2007bq}.

Apparently, the only different result obtained here is the importance
of the $P_{11}(1710)$ state. Most recent 
studies~\cite{Julia-Diaz:2006is,Mart:2006dk,Anisovich:2007bq,Arndt:2003ga}
found that this resonance has a negligible effect on the
$\gamma p \to K^+\Lambda$ reaction, in spite of the fact that
this resonance has long been used in the
isobar models of kaon photoproduction~\cite{previous_isobar}.
Although its contribution could be smaller than those of the
$S_{11}(1650)$, $P_{13}(1720)$, and $P_{13}(1900)$ resonances,
Fig.~\ref{fig:res_cont} recommends that a further detailed study
of the $C_x$ and $C_z$ data is urgently required to shed more
light on the role of the $P_{11}(1710)$ state.

Finally, we should mention that due to the nature of the prediction
here, experimental measurement of the $\sigma_{\rm TT'}$ is urgently
required to verify the current findings. Since the SAPHIR detector
has been dismantled, measurements of the $\sigma_{\rm TT'}$ by 
the CLAS, LEPS, and MAMI collaborations seem to be the only choice.

\section{\Large\sc Conclusion}
\label{conclusion}
In conclusion, we have calculated the contribution of the
kaon photoproduction $\gamma p\to K^+\Lambda$ channel to the
GDH sum rule by means of multipole models which fit the new
SAPHIR or CLAS data. Our findings indicate that the SAPHIR 
and CLAS data yield very different contributions to the GDH
sum rule. 
Contribution of this channel to the forward spin 
polarizability of the proton recommends a further analysis 
of the contribution from the $n\pi+\eta$ channels. 
The inclusion of the recent CLAS  $C_x$ and $C_z$ data
does not dramatically change this conclusion. The result of the fits 
that include these data supports the previous finding
that the $S_{11}(1650)$, $P_{13}(1720)$, 
and $P_{13}(1900)$ resonances
are essential for the $K^+\Lambda$ photoproduction. 
Future measurements of the $\sigma_{\rm TT'}$ by the
CLAS, LEPS, and MAMI collaborations, are expected to confirm 
this conclusion.

The author acknowledges the support from the University of
Indonesia.

\end{document}